\documentclass[aps,showpacs,twocolumn]{revtex4}
\usepackage{graphicx}

\begin{document}

\title{Solar system tests of Ho\v{r}ava-Lifshitz gravity}
\author{Tiberiu Harko}
\email{harko@hkucc.hku.hk}
\affiliation{Department of Physics and Center for Theoretical
and Computational Physics,
The University of Hong Kong, Pok Fu Lam Road, Hong Kong}
\author{Zoltan Kov\'{a}cs}
\email{zkovacs@mpifr-bonn.mpg.de}
\affiliation{Department of Physics and Center for Theoretical
and Computational Physics,
The University of Hong Kong, Pok Fu Lam Road, Hong Kong}
\author{Francisco S. N. Lobo}
\email{flobo@cii.fc.ul.pt} \affiliation{Centro de F\'{i}sica
Te\'{o}rica e Computacional, Faculdade de Ci\^{e}ncias da
Universidade de Lisboa, Avenida Professor Gama Pinto 2, P-1649-003
Lisboa, Portugal}
\date{\today}

\begin{abstract}

 In the present paper
we consider the possibility of observationally constraining Ho\v{r}ava
gravity at the scale of the Solar System, by considering the
classical tests of general relativity (perihelion precession of
the planet Mercury, deflection of light by the Sun and the radar
echo delay) for the spherically symmetric black hole Kehagias-Sfetsos solution of
Ho\v{r}ava-Lifshitz gravity. All these gravitational effects can
be fully explained in the framework of the vacuum solution of
Ho\v{r}ava gravity. Moreover, the study of the classical general
relativistic tests also constrain the free parameter of the
solution. From the analysis of the perihelion precession of the planet Mercury we obtain for the free parameter $\omega $ of the Kehagias-Sfetsos solution the constraint $\omega \geq 3.212\times 10^{-26}\;{\rm cm}^{-2}$, the deflection of light by the Sun gives $\omega \geq 4.589\times 10^{-26}\;{\rm cm}^{-2}$, while the radar echo delay observations can be explained if the value of $\omega $ satisfies the constraint $\omega \geq 9.179\times 10^{-26}\;{\rm cm}^{-2}$.

\end{abstract}

\maketitle


\section{Introduction}

Recently, a renormalizable gravity theory in four dimensions which
reduces to Einstein gravity with a non-vanishing cosmological
constant in the infrared (IR) energy scale (corresponding to large distances), but with improved ultraviolet (UV) energy scale behaviors (corresponding to very small distances), was proposed by
Ho\v{r}ava \cite {Horava:2008ih,Horava:2009uw}. Quantum field theory has had considerable success experimentally, but from a theoretical point of view it predicts infinite values for physical quantities. Thus for certain Feynman diagrams containing loops, the calculations leads to an infinite result. This is known as the ultraviolet divergence of quantum field theory, because small loop sizes correspond to high energies. Infra-red Divergence
is the divergence  caused by the low energy behavior of a quantum theory. The  Ho\v{r}ava theory
admits a Lifshitz scale-invariance in time and space, exhibiting a
broken Lorentz symmetry at short scales, while at large distances
higher derivative terms do not contribute, and the theory reduces
to standard general relativity (GR). The Ho\v{r}ava theory has
received a great deal of attention and since its formulation
various properties and characteristics have been extensively
analyzed, ranging from formal developments \cite{formal},
cosmology \cite{cosmology}, dark energy \cite{darkenergy} and dark
matter \cite{darkmatter}, spherically symmetric or rotating
solutions
\cite{BHsolutions,Park:2009zra,Lu:2009em,Kehagias:2009is, ghodsi}, to the weak field observational tests \cite{iorio,iorio1}.
At large distances, higher derivative terms do not contribute and the theory reduces to standard
general relativity if a particular coupling $\lambda $, which controls the contribution of the trace of the extrinsic curvature has a specific value. Indeed, $\lambda $ is running, and if $\lambda = 1$ is an IR fixed point, standard
general relativity is recovered. Therefore, although a generic vacuum of the theory is the anti-de Sitter one,
particular limits of the theory allow for the Minkowski vacuum, a physical state characterized by the absence of ordinary (baryonic) matter. In
this limit post-Newtonian coefficients coincide with those of pure
GR. Thus, the deviations from conventional GR can be tested only
beyond the post-Newtonian corrections, that is for a system with
strong gravity at astrophysical scales.

In this context, IR-modified Ho\v{r}ava gravity seems to be
consistent with the current observational data, but in order to
test its viability more observational constraints are necessary.
In Ref.~\cite{Konoplya:2009ig}, potentially observable properties
of black holes in the Ho\v{r}ava-Lifshitz gravity with Minkowski
vacuum were considered, namely, the gravitational lensing and
quasinormal modes. Quasinormal modes are the modes of energy dissipation of a perturbed object or field. Black holes have many quasinormal modes that describe the exponential decrease of asymmetry of the black hole in time as it evolves towards the spherical shape. It was shown that the bending angle is
seemingly smaller in the considered Ho\v{r}ava-Lifshitz gravity
than in GR, and the quasinormal modes of black holes are longer
lived, and have larger real oscillation frequency in
Ho\v{r}ava-Lifshitz gravity than in GR. In
Ref.~\cite{Chen:2009eu}, by adopting the strong field limit
approach, the properties of strong field gravitational lensing for
the H\v{o}rava-Lifshitz black hole were considered, and
the angular position and magnification of the relativistic images
were obtained. Compared with the Reissner-Norstr\"{o}m black hole,
a significant difference in the parameters was found. Thus, it was
argued that this may offer a way to distinguish a deformed
H\v{o}rava-Lifshitz black hole from a Reissner-Nordstr\"{o}m black
hole. In general relativity the Reissner-Nordstr\"{o}m black
hole solution describes the gravitational field of a charged black hole. In Ref.~\cite{Chen:2009bu}, the behavior of the effective
potential was analyzed, and the timelike geodesic motion in the
Ho\v{r}ava-Lifshitz spacetime was also explored.
In Ref.~\cite{HKL-accretion}, the basic physical properties of
matter forming a thin accretion disk in the vacuum spacetime
metric of the Ho\v{r}ava-Lifshitz gravity models were considered.
It was shown that significant differences as compared to the
general relativistic case exist, and that the determination of
these observational quantities could discriminate, at least in
principle, between standard GR and Ho\v{r}ava-Lifshitz theory, and
constrain the parameter of the model.

It is the purpose of the present paper to consider the classical
tests (perihelion precession, light bending, and the radar echo
delay, respectively) of general relativity for static
gravitational fields in the framework of Ho\v{r}ava-Lifshitz
gravity. To do this we shall adopt for the geometry outside a
compact, stellar type object (the Sun), the static and spherically
symmetric metric obtained by Kehagias and Sfetsos
\cite{Kehagias:2009is}. For the Kehagias and Sfetsos (KS) metric,
we first consider the motion of a particle (planet), and analyze
the perihelion precession. In addition to this, by considering the
motion of a photon in the static KS field, we study the bending of
light by massive astrophysical objects and the radar echo delay,
respectively. All these gravitational effects can be explained in
the framework of the KS geometry. Existing data on light-bending
around the Sun, using long-baseline radio interferometry, ranging
to Mars using the Viking lander, the perihelion precession of
Mercury, and recent Lunar Laser Ranging results can all give
significant and detectable Solar System constraints, associated
with Ho\v{r}ava-Lifshitz gravity. More exactly, the study of the
classical general relativistic tests, constrain the parameter of
the solution. We will also compare our results with the phenomenological constraints
on the Kehagias-Sfetsos solution from solar system orbital motions obtained
in \cite{iorio}. In order to obtain reliable constraints on the parameter of the model, the analysis must not be limited to the perihelion precession of the planet Mercury, but it should be extended to take into account the other recent observational data on the perihelion precession of the other planets of the Solar System.

In the context of the classical tests of GR the Dadhich, Maartens,
Papadopoulos and Rezania (DMPR) solution of the spherically
symmetric static vacuum field equations in brane world models was
also extensively analyzed \cite{Boehmer:2008zh}. It was found that
the existing observational solar system data constrain the
numerical values of the bulk tidal parameter and of the brane
tension.

This paper is organized as follows. In Sec.~\ref{sec:II}, we
present the action and specific solutions of static and
spherically symmetric spacetimes in Ho\v{r}ava-Lifshitz gravity.
In Sec.~\ref{sect4}, we analyze the classical Solar System tests
for the case of the KS asymptotically flat solution
\cite{Kehagias:2009is} of Ho\v{r}ava-Lifshitz gravity. We conclude
our results in Sec.~\ref{sect5}.

\section{Black holes in Ho\v{r}ava-Lifshitz gravity}
\label{sec:II}

In the Ho\v{r}ava-Lifshitz gravity theory, Lorentz symmetry is broken in the
ultraviolet. The breaking manifests in the strong
anisotropic scalings between space and time,
$\vec{x}\rightarrow l\vec{x}$, $t\rightarrow l^zt$.
In (3 + 1)-dimensional spacetimes, the theory is powercounting
renormalizable, provided that $z\geq 3$. At low
energies, the theory is expected to flow to $z = 1$, whereby
the Lorentz invariance is accidentally restored. Such an
anisotropy between time and space can be easily realized,
when writing the metric in the Arnowitt-Deser-Misner
(ADM) form \cite{MWT}. The formalism supposes that the spacetime is foliated into a family of spacelike surfaces $\Sigma _t$, labeled by their time coordinate $t$, and with coordinates on each slice given by $x^i$. Using the ADM formalism, the four-dimensional metric is parameterized in the following form
\begin{equation}
ds^{2}=-N^{2}c^{2}dt^{2}+g_{ij}\left( dx^{i}+N^{i}\,dt\right) \left(
dx^{j}+N^{j}\,dt\right) .
\end{equation}
The Einstein-Hilbert action is given by
\begin{equation}
S=\frac{1}{16\pi G}\int d^{4}x\;\sqrt{g}\,N\left(
K_{ij}K^{ij}-K^{2}+R^{(3)}-2\Lambda \right) ,  \label{EHaction}
\end{equation}
where $G$ is Newton's constant, $R^{(3)}$ is the three-dimensional
curvature scalar for $g_{ij}$, and $K_{ij}$ is the extrinsic
curvature, defined as
\begin{equation}
K_{ij}=\frac{1}{2N}\left( \dot{g}_{ij}-\nabla _{i}N_{j}-\nabla
_{j}N_{i}\right) ,
\end{equation}
where the dot denotes a derivative with respect to $t$.

The IR-modified Ho\v{r}ava action is given by \cite{Park:2009zra}
\begin{eqnarray}
S &=&\int dt\,d^{3}x\;\sqrt{g}\,N\Bigg[\frac{2}{\kappa ^{2}}\left(
K_{ij}K^{ij}-\lambda _{g}K^{2}\right) -\frac{\kappa ^{2}}{2\nu
_{g}^{4}}
C_{ij}C^{ij}
   \nonumber \\
&&+\frac{\kappa ^{2}\mu }{2\nu _{g}^{2}}\epsilon ^{ijk}R_{il}^{(3)}\nabla
_{j}R^{(3)l}{}_{k}-\frac{\kappa ^{2}\mu ^{2}}{8}R_{ij}^{(3)}R^{(3)ij}
\nonumber \\
&&+\frac{\kappa ^{2}\mu ^{2}}{8(3\lambda _{g}-1)}\left(
\frac{4\lambda _{g}-1
}{4}(R^{(3)})^{2}-\Lambda _{W}R^{(3)}+3\Lambda _{W}^{2}\right)
   \nonumber \\
&&+\frac{\kappa ^{2}\mu ^{2}\omega }{8(3\lambda _{g}-1)}R^{(3)}\Bigg],
\label{Haction}
\end{eqnarray}
where $\kappa $, $\lambda _{g}$, $\nu _{g}$, $\mu $, $\omega $ and
$\Lambda _{W}$ are constant parameters. $K=K_i^i$ is the contraction of the intrinsic curvature, while $\nabla
_{j}$ is the covariant derivative with respect to the three-dimensional metric. $C^{ij}$ is the Cotton
tensor, defined as
\begin{equation}
C^{ij}=\epsilon ^{ikl}\nabla _{k}\left( R^{(3)j}{}_{l}-\frac{1}{4}
R^{(3)}\delta _{l}^{j}\right) .
\end{equation}
In the UV, this theory is power counting renormalizable, at least around the flat space (vacuum)
solution. In the IR, the terms of lowest dimension should dominate. Note that the last term in Eq.~(\ref{Haction}) represents a `soft' violation of the `detailed balance' condition, which modifies the
IR behavior. More specifically, if one maintains Ho\v{r}ava's original detailed balance, and try to recover
Einstein-Hilbert in the low energy regime, then one obtains a negative cosmological constant. The introduction of the last term in Eq.~(\ref{Haction}) corrects this, and provides a positive cosmological constant. Since $\omega \propto \mu ^2$, this IR modification term, $\mu ^{4}R^{(3)}$, with an arbitrary
cosmological constant, represents the analogs of the standard
Schwarzschild-(A)dS solutions, which were absent in the original
Ho\v{r}ava model.

The fundamental constants of the speed of light $c$, Newton's
constant $G$, and the cosmological constant $\Lambda$ are defined
as
\begin{equation}
c^2=\frac{\kappa^2\mu^2|\Lambda_W|}{8(3\lambda_g-1)^2}\quad
G=\frac{ \kappa^2c^2}{16\pi(3\lambda_g-1)}\quad
\Lambda=\frac{3}{2}\Lambda_W c^2.
\end{equation}

The Ho\v{r}ava-Lifshitz theory is associated with the breaking of the diffeomorphism invariance, required for the anisotropic scaling in the UV \cite{Horava:2009uw}.

Throughout this work, we consider the static and spherically
symmetric metric given by
\begin{equation}
ds^{2}=-e^{\nu (r)}dt^{2}+e^{\lambda (r)}dr^{2}+r^{2}\,(d\theta
^{2} +\sin^{2}{\theta }\,d\phi ^{2}),  \label{SSSmetric}
\end{equation}
where $e^{\nu (r)}$ and $e^{\lambda (r)}$ are arbitrary functions
of the radial coordinate $r$.

Imposing the specific case of $\lambda _g=1$, which reduces to the
Einstein-Hilbert action in the IR limit, one obtains the following
solution of the vacuum field equations in Ho\v{r}ava gravity,
\begin{equation}
e^{\nu (r)}=e^{-\lambda (r)}=1+(\omega-\Lambda_W)r^2-\sqrt{
r[\omega(\omega-2\Lambda_W)r^3 +\beta]},  \label{gensolution}
\end{equation}
where $\beta$ is an integration constant \cite{Park:2009zra}.

By considering $\beta=-\alpha^2/\Lambda_W$ and $\omega=0$ the
solution given by Eq.~(\ref{gensolution}) reduces to the Lu, Mei
and Pope (LMP) solution \cite{Lu:2009em}, given by
\begin{equation}
e^{\nu
(r)}=1-\Lambda_Wr^2-\frac{\alpha}{\sqrt{-\Lambda_W}}\sqrt{r}.
\label{LMPsolution}
\end{equation}

Alternatively, considering now $\beta=4\omega M$ and
$\Lambda_W=0$, one obtains the Kehagias and Sfetsos's (KS)
asymptotically flat solution \cite {Kehagias:2009is}, given by
\begin{equation}
e^{\nu (r)}=1+\omega r^2-\omega r^2\sqrt{1+\frac{4M}{\omega r^3}}.
\label{KSsolution}
\end{equation}
If the limit $4M/\omega r^3\ll 1$, from Eq.~(\ref{KSsolution}) we
obtain the standard Schwarzschild metric of general relativity,
$e^{\nu (r)}=1-2M/r$, which represents a ``Post-Newtonian''
approximation of the KS solution of the second order in the speed
of light.  We shall use the KS solution for analyzing the Solar
System constraints of the theory. Note that there are two event
horizons at
\begin{equation}
r_{\pm}=M\left[1\pm\sqrt{1-1/(2\omega M^2)}\right].
\end{equation}
To avoid a naked singularity at the origin, one also needs to
impose the condition
\begin{equation}
\omega M^2\geq \frac{1}{2}.
\end{equation}
Note that in the GR regime, i.e., $\omega M^2 \gg 1$, the outer
horizon approaches the Schwarzschild horizon, $r_+\simeq 2M$, and
the inner horizon approaches the central singularity, $r_- \simeq
0$. One should also note that the KS solution is obtained without requiring the projectability condition, which was assumed in the original Ho\v{r}ava theory. However, it is important to emphasize that static and spherically symmetric exhibiting the projectability condition have been obtained in the literature \cite{project-cond}.

\section{Solar system tests for Ho\v{r}ava-Lifshitz gravity black holes}
\label{sect4}

At the level of the Solar System there are three fundamental
tests, which can provide important observational evidence for GR
and its generalizations, and for alternative theories of
gravitation in flat space. These tests are the precession of the
perihelion of Mercury, the deflection of light by the Sun, and the
radar echo delay observations, respectively, and have been used to
successfully test the Schwarzschild solution of general relativity
and some of its generalizations. In this Section we consider these
standard Solar System tests of general relativity in the case of
the KS asymptotically flat solution \cite{Kehagias:2009is} of
Ho\v{r}ava-Lifshitz gravity. Throughout the next Sections we use
the natural system of units with $G=c=1$.

\subsection{The perihelion precession of the planet Mercury}

The motion of a test particle in the gravitational field of the
metric given by Eq.~(\ref{SSSmetric}) can be derived from the
variational principle
\begin{equation}
\delta \int \sqrt{-e^{\nu }\dot{t}^{2}+e^{\lambda }\dot{r}^{2}
+r^{2}\left( \dot{\theta}^{2}+\sin ^{2}\theta
\dot{\phi}^{2}\right) }ds=0, \label{var}
\end{equation}
where the dot denotes $d/ds$. It may be verified that the orbit is
planar, and hence without any loss of generality we can set
$\theta =\pi /2$. Therefore we will use $\phi $ as the angular
coordinate. Since neither $t$ nor $\phi $ appear explicitly in Eq.
(\ref{var}), their conjugate momenta are constant,
\begin{equation}
e^{\nu }\dot{t}=E=\mathrm{constant}, \qquad
r^{2}\dot{\phi}=L=\mathrm{constant}. \label{consts}
\end{equation}

The line element, given by Eq.~(\ref{SSSmetric}), and taking into
account Eqs. (\ref{consts}), provides the following equation of
motion for $r$
\begin{equation}
\dot{r}^{2}+e^{-\lambda }\frac{L^{2}}{r^{2}}=e^{-\lambda }\left(
E^{2}e^{-\nu }-1\right) .  \label{inter1}
\end{equation}

The change of variable $r=1/u$ and the substitution
$d/ds=Lu^{2}d/d\phi $ transforms Eq.~(\ref{inter1}) into the form
\begin{equation}
\left( \frac{du}{d\phi }\right) ^{2}+e^{-\lambda
}u^{2}=\frac{1}{L^{2}} e^{-\lambda }\left( E^{2}e^{-\nu }-1\right)
.
\end{equation}
By formally representing $e^{-\lambda }=1-f(u)$, we obtain
\begin{equation}
\left( \frac{du}{d\phi }\right)
^{2}+u^{2}=f(u)u^{2}+\frac{E^{2}}{L^{2}} e^{-\nu -\lambda
}-\frac{1}{L^{2}}e^{-\lambda }. \label{ueq_basic}
\end{equation}
By taking the derivative of the previous equation with respect to
$\phi $ we find
\begin{equation}
\frac{d^{2}u}{d\phi ^{2}}+u=F(u),  \label{inter2}
\end{equation}
where
\begin{equation}
F(u)=\frac{1}{2}\frac{dG(u)}{du},
\end{equation}
and we have denoted
\[
G(u)\equiv f(u)u^{2}+\frac{E^{2}}{L^{2}} e^{-\nu -\lambda
}-\frac{1}{L^{2}}e^{-\lambda }.
\]

A circular orbit $u=u_{0}$ is given by the root of the equation $
u_{0}=F\left( u_{0}\right) $. Any deviation $\delta =u-u_{0}$ from
a circular orbit must satisfy the equation
\begin{equation}
\frac{d^{2}\delta }{d\phi ^{2}}+\left[ 1-\left(
\frac{dF}{du}\right) _{u=u_{0}}\right] \delta =O\left( \delta
^{2}\right) ,
\end{equation}
which is obtained by substituting $u=u_{0}+\delta $ into Eq.
(\ref{inter2}). Therefore, in the first order in $\delta $, the
trajectory is given by
\begin{equation}
\delta =\delta _{0}\cos \left( \sqrt{1-\left( \frac{dF}{du}\right)
_{u=u_{0}} }\phi +\beta \right) ,
\end{equation}
where $\delta _{0}$ and $\beta $ are constants of integration. The
angles of the perihelia of the orbit are the angles for which $r$
is minimum, and hence $u$ or $\delta $ is maximum. Therefore, the
variation of the orbital angle from one perihelion to the next is
\begin{equation}
\phi =\frac{2\pi }{\sqrt{1-\left( \frac{dF}{du}\right)
_{u=u_{0}}}}=\frac{ 2\pi }{1-\sigma }.  \label{prec}
\end{equation}

The quantity $\sigma $ defined by the above equation is called the
perihelion advance, which represents the rate of advance of the
perihelion. As the planet advances through $\phi $ radians in its
orbit, its perihelion advances through $\sigma \phi $ radians.
From Eq. (\ref{prec}) $\sigma $ is given by
\begin{equation}
\sigma =1-\sqrt{1-\left( \frac{dF}{du}\right) _{u=u_{0}}},
\end{equation}
or, for small $\left( dF/du\right) _{u=u_{0}}$, by
\begin{equation}
\sigma =\frac{1}{2}\left( \frac{dF}{du}\right) _{u=u_{0}}.
\end{equation}

For a complete rotation we have $\phi \approx 2\pi (1+\sigma )$,
and the advance of the perihelion is
\begin{equation}
\delta \phi =\phi -2\pi \approx 2\pi \sigma .
\end{equation}

The observed value of the perihelion precession of the planet
Mercury is $ \delta \phi _{Obs}=43.11\pm 0.21$ arcsec per century
\cite{Sh}. As a first step in the study of the perihelion
precession in Ho\v{r}ava-Lifshitz gravity, the relevant functions
are given by
\[
f(u)=-\frac{\omega}{u^{2}}+\frac{\omega}{u^{2}} \sqrt{
1+\frac{4Mu^{3}}{\omega} }\,,
\]
\begin{eqnarray}
G(u) &=&-\omega \left( 1+\frac{1}{L^{2}u^{2}}\right) \left(
1-\sqrt{1+\frac{
4M}{\omega }u^{3}}\right) +  \nonumber \\
&&\frac{1}{L^{2}}\left( E^{2}-1\right) \nonumber,
\end{eqnarray}
and
\begin{eqnarray}
F(u)&=&\frac{\omega }{L^{2}u^{3}}\left( 1-\sqrt{1+\frac{4M}{\omega
}u^{3}}
\right) +  \nonumber \\
&&3M\left( 1+\frac{1}{L^{2}u^{2}}\right)
\frac{u^{2}}{\sqrt{1+\frac{4M}{ \omega }u^{3}}}\,,
\end{eqnarray}
respectively.

The circular orbits are given by the roots of the equation $F\left(
u_{0}\right) =u_{0}$, which is given by
\begin{equation}
3Mu_{0}^{2}-\frac{M}{L^{2}}=\frac{\omega
}{L^{2}u_{0}^{3}}+\sqrt{1+\frac{4M}{ \omega }u_{0}^{3}}\left(
u_{0}-\frac{\omega }{L^{2}u_{0}^{3}}\right) . \label{eq1}
\end{equation}

In order to solve Eq.~(\ref{eq1}) we represent
the parameter $\omega $ as $%
\omega =\omega _{0}/M^{2}$, and $u_{0}$ as $u_{0}=x_{0}/M$, where
$\omega _{0}$ and $x_{0}$ are dimensionless parameters,
respectively. Then Eq.~(\ref {eq1}) can be written as
\begin{equation}
3x_{0}^{2}-b^{2}=\frac{\omega _{0}b^{2}}{x_{0}^{3}}
+\sqrt{1+\frac{4}{\omega_{0}}x_{0}^{3}}\left( x_{0}-\frac{\omega
_{0}b^{2}}{x_{0}^{3}}\right) , \label{eqx}
\end{equation}
where $b^{2}=M^{2}/L^{2}$.

\begin{figure}[h]
\includegraphics[width=3.05in]{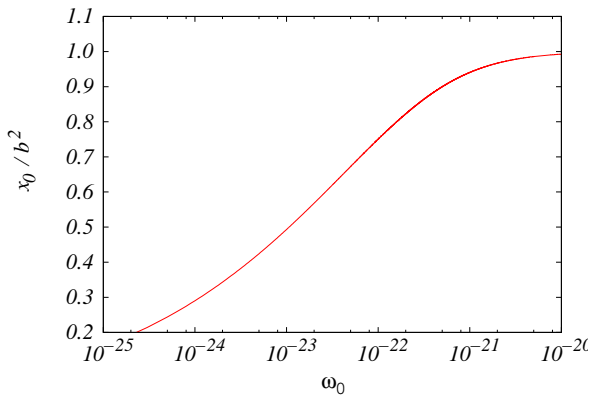}
\caption{Variation of $x_0$ as a function of $\omega _0$.}
\label{fig1}
\end{figure}

In the case of the planet Mercury we have $a=57.91\times 10^{11}$
cm and $ e=0.205615$, while for the values of the mass of the Sun
and of the physical constants we take $M=M_{\odot }=1.989\times
10^{33}$ g, $c=2.998\times 10^{10}$ cm/s, and $G=6.67\times
10^{-8}$ cm$^{3}$g$^{-1}$s$^{-2}$, respectively \cite{An}. Mercury
also completes $415.2$ revolutions each century. With the use of
these numerical values we first obtain $ b^{2}=M/a\left(
1-e^{2}\right) =2.66136\times 10^{-8}$. By performing a first
order series expansion of the square root in Eq.~(\ref{eqx}), we
obtain the standard general relativistic equation
$3x_{0}^{2}-x_{0}+b^{2}=0$, with the physical solution
$x_{0}^{(GR)}\approx b^{2}$. In the general case, the value of
$x_{0}$ also depends on the numerical value of $\omega _{0}$, and,
for a given $\omega _{0}$, $x_{0}$ must be obtained by numerically
solving the nonlinear algebraic equation Eq.~(\ref{eqx}). The
variation of $x_{0}$ as a function of $\omega _{0}$ is represented
in Fig.~\ref{fig1}.

The perihelion precession is given by $\delta \phi =\pi
\left(dF(u)/du\right)|_{u=u_0}$, and in the variables $x_0$ and
$\omega _0$ can be written as
\begin{equation}
\delta \phi =\pi\frac{3 \sqrt{\omega _0 } \Phi\left(\omega _0,x_0\right)}{x^4_0 \left(4
   x^3_0+\omega _0 \right)^{3/2}},
\end{equation}
where
\begin{eqnarray}
\Phi \left(\omega _0, x_0\right)&=&2
\left(x^3_0+\omega _0 \right) x^5_0+\nonumber\\
&&b^2 \Big[2 x^6_0
+\left(6 \omega _0 -4 \sqrt{\omega _0  \left(4
   x^3_0+\omega _0 \right)}\right) x^3_0
   \nonumber  \\
   &&+\omega _0 ^2
   -\sqrt{\omega _0 ^3 \left(4 x^3_0+\omega _0
   \right)}\Big].
\end{eqnarray}

\begin{figure}[h]
\includegraphics[width=3.05in]{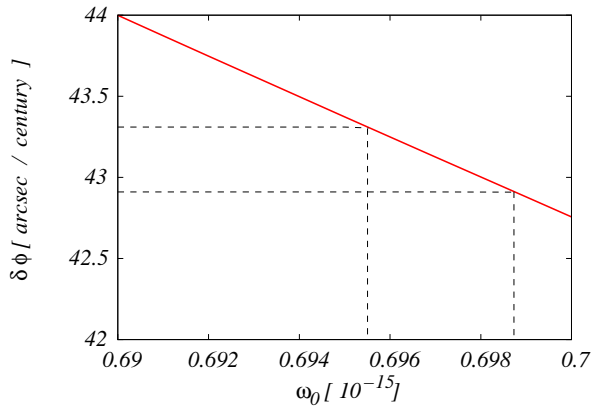}
\caption{Variation of the planetary precession angle $\delta \phi$
(in arcseconds per century) as a function of $\omega _0$.}
\label{fig2}
\end{figure}

In the ``Post-Newtonian'' limit $4x_0^3/\omega _0\ll 1$, we obtain
the classical general relativistic result $\delta \phi _{GR}=6\pi
b^2$. The variation of the perihelion precession angle as a
function of $\omega _0$ is represented in Fig.~\ref{fig2}.

The gravity analysis of radio Doppler and range data generated by
the Deep Space Network with Mariner 10 during two of its
encounters with Mercury in March 1974 and March 1975 determined
the observed value of the perihelion precession of the planet
Mercury as $\delta \phi _{Obs}=43.11\pm 0.20$ arcsec per century
\cite{An}. Therefore the range of variation of the perihelion
precession is $\delta \phi _{Obs}\in \left(42.91,43.31\right)$.
This range of observational values fixes the range of $\omega _0$
as $\omega _0\in\left(6.95508\times 10^{-16},6.98748\times
10^{-16}\right)$. The general relativistic formula for the
precession, gives $\delta \phi _{GR}=42.94$ arcsec per century.
The perihelion precession can be also obtained by using elliptic
integrals, or, in the case of the arbitrary central potentials, by
using the method developed in \cite{Schmidt}.

Recently, new observational results on the extra-precession of the perihelion of the planets of the Solar System have been obtained. These results can be used as a test for gravitational models in the Solar System, like, for example, the determinations of the PPN parameter $\beta $ and of the coefficient $J_2$ of the oblateness of the Sun \cite{pit}. One can also use these results to estimate a possible advance in the planets perihelia, and an anomalous precession to the usual Newtonian/Einsteinian secular precession of the longitude of the perihelion of Saturn was found \cite{iorio2}. Once the precision  of the observations would improve, these data could also be used to constrain through perihelion precession the value of the parameter $\omega $ for the Kehagias-Sfetsos black hole solution.

\subsection{Light deflection by the Sun}

The deflection angle of light rays passing nearby the Sun in the
KS geometry is given, with the use of general relation derived in
\cite{Wein}, by
\begin{equation}
\phi \left( r\right) =\phi \left( \infty \right) +\int_{r}^{\infty }\frac{%
\left[f(r)\right]^{-1/2}}{\sqrt{f\left( r_{0}\right) \left( r/r_{0}\right) ^{2}/f(r)-1}}%
\frac{dr}{r},  \label{defl4}
\end{equation}
where $r_{0}$ is the distance of
the closest approach, and we have denoted
\begin{equation}
f(r)=1+\omega r^{2}-\sqrt{r(\omega ^{2}r^{3}+4\omega M)}.
\end{equation}
By introducing a new variable $x$ by means of the transformation
$r=r_{0}x$, Eq.~(\ref{defl4}) can be written as
\begin{equation}
\phi \left( r_{0}\right) =\phi \left( \infty \right) +\int_{1}^{\infty }%
\frac{\left[f\left( r_{0}x\right)\right]^{-1/2} }{\sqrt{f\left( r_{0}\right)
x^{2}/f\left( r_{0}x\right) -1}}\frac{dx}{x}
\end{equation}

By representing $\omega $ as $\omega =\omega _{0}/M^{2}$, and
$r_{0}$ as $ r_{0}=x_{0}M$, we obtain
\begin{equation}
\phi \left( x_{0}\right) =\phi \left( \infty \right) +\int_{1}^{\infty }%
\frac{\left[ g\left( \omega _{0},x_{0},x\right) \right] ^{-1/2}}{\sqrt{%
g_{0}\left( \omega _{0},x_{0}\right) x^{2}/g\left( \omega
_{0},x_{0},x\right) -1}}\frac{dx}{x},
\end{equation}
where we have denoted
\begin{equation}
g\left( \omega _{0},x_{0},x\right) =1+\omega _{0}x_{0}^{2}x^{2}-\sqrt{%
x_{0}x(\omega _{0}^{2}x_{0}^{3}x^{3}+4\omega _{0})},
\end{equation}
and
\begin{equation}
g_{0}\left( \omega _{0},x_{0}\right) =1+\omega _{0}x_{0}^{2}-\sqrt{%
x_{0}(\omega _{0}^{2}x_{0}^{3}+4\omega _{0})},
\end{equation}
respectively.

\begin{figure}[tbp]
\includegraphics[width=3.05in]{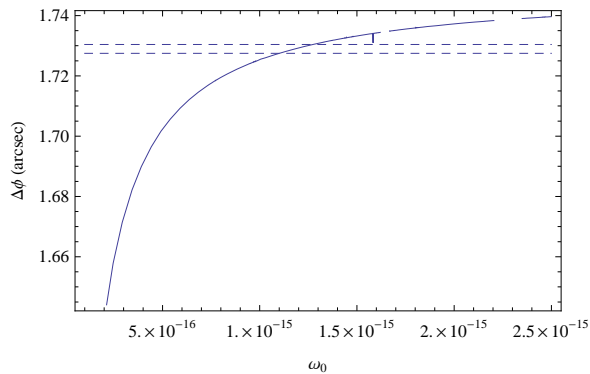}
\caption{The light deflection angle $\Delta \phi $ (in arcseconds)
as a function of the parameter $\omega _0$.} \label{fig3}
\end{figure}

For the Sun, by taking $r_{0}=R_{\odot }=6.955\times 10^{10}$ cm,
where $R_{\odot}$ is the radius of the Sun, we find for $x_{0}$
the value $x_{0}=4.71194\times 10^{5}$. The variation of the
deflection angle $\Delta \phi =2\left| \phi \left( x_{0}\right)
-\phi \left( \infty \right) \right| -\pi $ is represented, as a
function of $\omega _{0}$, in Fig.~\ref{fig3}. In the "Post-Newtonian" limit $4x_0^3/\omega _0\ll 1$, we obtain
the classical general relativistic result $\Delta \phi=4M_{\odot}/R_{\odot}=1.73''$.

We consider now the constraints on the Ho\v{r}ava-Lifshitz gravity
arising from the solar system observations of the light deflection
by the Sun. The best available data come from long baseline radio
interferometry \cite{all2}, which gives $\delta \phi _{LD}=\delta
\phi _{LD}^{(GR)}\left( 1+\Delta _{LD}\right) $, with $\Delta
_{LD}\leq 0.0017$, where $\delta \phi _{LD}^{(GR)}=1.7275 $
arcsec. The observational constraints of light deflection
restricts the value of $\omega _0$ to  $\omega _0\in
\left(1.1\times10^{-15},1.3\times 10^{-15}\right)$.

\subsection{Radar echo delay}

A third Solar System test of general relativity is the radar echo
delay \cite{Sh}. The idea of this test is to measure the time
required for radar signals to travel to an inner planet or
satellite in two circumstances: a) when the signal passes very
near the Sun and b) when the ray does not go near the Sun. The
time of travel of light $t_{0}$ between two planets, situated far
away from the Sun, is given by
\begin{equation}
t_{0}=\int_{-l_{1}}^{l_{2}}dy,
\end{equation}
where $dy$ is the differential distance in the radial direction,
and $l_{1}$ and $l_{2}$ are the distances of the planets to the
Sun, respectively. If the light travels close to the Sun, in the
metric given by Eq.~(\ref{SSSmetric}) the time travel is
\begin{equation}
t=\int_{-l_{1}}^{l_{2}}\frac{dy}{v}=\int_{-l_{1}}^{l_{2}}e^{\left[\lambda
(r)-\nu (r)\right] /2}dy,
\end{equation}
 $v=e^{\left( \nu -\lambda \right) /2}$ is the speed of light
in the presence of the gravitational field. The time difference is
\begin{equation}
\Delta t=t-t_{0}=\int_{-l_{1}}^{l_{2}}\left\{ e^{\left[ \lambda
(r) -\nu (r) \right] /2}-1\right\} dy.
\end{equation}

Since $r=\sqrt{y^{2}+R_{\odot }^{2}}$, where $R_{\odot }$ is the
radius of the Sun, we have
\begin{equation}  \label{echo}
\Delta t=\int_{-l_{1}}^{l_{2}}\left\{ e^{\left[ \lambda \left(
\sqrt{ y^{2}+R_{\odot }^{2}}\right) -\nu \left(
\sqrt{y^{2}+R_{\odot }^{2}}\right) \right] /2}-1\right\} dy.
\label{delay_eq}
\end{equation}

The first experimental Solar System constraints on time delay
have come from the Viking lander on Mars \cite{Sh}. In the Viking
mission two transponders landed on Mars and two others continued
to orbit round it. The latter two transmitted two distinct bands
of frequencies, and thus the Solar coronal effect could be
corrected for. However, recently the measurements of the frequency
shift of radio photons to and from the Cassini spacecraft as they
passed near the Sun have greatly improved the observational
constraints on the radio echo delay. For the time delay of the
signals emitted on Earth, and which graze the Sun, one obtains
$\Delta t_{RD}=\Delta t_{RD}^{(GR)}\left( 1+\Delta _{RD}\right) $,
with $\Delta _{RD}\leq \left(2.1\pm2.3\right)\times 10^{-5}$
\cite{Re79}.

\begin{figure}[tbp]
\includegraphics[width=3.05in]{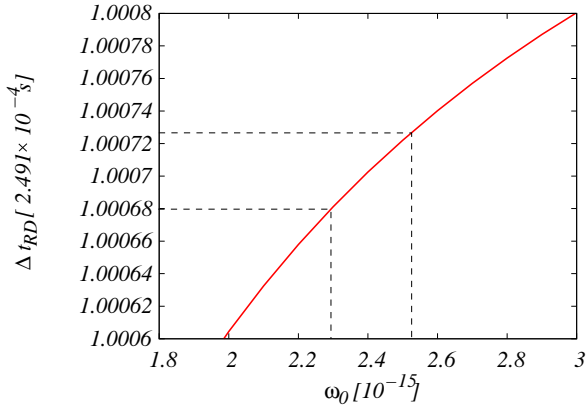}
\caption{Variation of the time delay $\Delta t_{RD}$ as a function
of $\omega _0$.} \label{fig4}
\end{figure}

For the case of the Earth-Mars-Sun system we have
$R_{E}=l_{1}=1.525\times 10^{13}\,\mathrm{cm}$ (the distance
Earth-Sun) and $R_{P}=l_{2}=2.491\times 10^{13}\,\mathrm{cm}$ (the
distance Mars-Sun).  With these values the standard general
relativistic radar echo delay has the value $\Delta
t_{RD}^{(GR)}\approx 4M_{\odot}\ln \left( 4l_{1}l_{2}/R_{\odot
}^{2}\right) \approx 2.4927\times 10^{-4}\, \mathrm{s}$. With the
use of Eq.~(\ref{echo}), it follows that the time delay for the KS
black hole solution of Ho\v{r}ava-Lifshitz gravity can be
represented as
\begin{widetext}
\begin{equation}
\Delta t_{RD}=2\int_{-l_{1}}^{l_{2}}\frac{\omega \left(
y^{2}+R_{\odot}^{2}\right) \left[ \sqrt{1+\left( 4M_{\odot}/\omega
\right) \left( y^{2}+R_{\odot}^{2}\right) ^{-3/2}}-1\right]
}{1-\omega \left( y^{2}+R_{\odot}^{2}\right) \left[ \sqrt{1+\left(
4M_{\odot}/\omega \right) \left( y^{2}+R_{\odot}^{2}\right)
^{-3/2}}-1 \right] }dy.
\end{equation}

By introducing a new variable $\xi $ defined as $y=2\xi
M_{\odot}$, and by representing again $\omega $ as $\omega =\omega
_{0}/M_{\odot}^{2}$, we obtain for the time delay, the following
expression
\begin{equation}
\Delta t_{RD}=16\omega _{0}M_{\odot}\int_{-\xi _{1}}^{\xi
_{2}}\frac{ \left( \xi ^{2}+a^{2}\right) \left[ \sqrt{1+\left(
1/2\omega _{0}\right) \left( \xi ^{2}+a^{2}\right)
^{-3/2}}-1\right] }{1-4\omega _{0}\left( \xi ^{2}+a^{2}\right)
\left[ \sqrt{1+\left( 1/2\omega _{0}\right) \left( \xi
^{2}+a^{2}\right) ^{-3/2}}-1\right] }d\xi ,
\end{equation}
\end{widetext}
where $a^{2}=R_{\odot }^{2}/4M_{\odot }^{2}$, $\xi
_{1}=l_{1}/2M_{\odot }$, and $\xi _{2}=l_{2}/2M_{\odot }$,
respectively. The variation of the time delay as a function of
$\omega _0$ is represented in Fig.~\ref{fig4}. In the limiting case $4M/\omega r^3\ll 1$ we reobtain again the standard general relativistic result \cite{Sh}.

The observational values of the radar echo delay are consistent
with the KS black hole solution in Ho\v{r}ava-Lifshitz gravity if
$\omega _0\in\left(2.0199\times 10^{-15},2.2000\times
10^{-15}\right)$. The general relativistic value $\Delta
t_{RD}=2.4927\times 10^{-4}$ is obtained for $\omega _0\approx
4\times 10^{-15}$.

\section{Discussions and final remarks}\label{sect5}

In the present paper, we have considered  observational possibilities
for testing the Kehagias and Sfetsos solution of the vacuum field equations in Ho\v{r}ava-Lifshitz gravity at the level of the Solar System. We have found that this
solution can give a  satisfactory description of the
perihelion precession of the planet Mercury, and of the other gravitational phenomena in the Solar System.  The classical tests
of general relativity (perihelion precession, light deflection and
radar echo delay) give strong constraints on the numerical value
of the parameter $\omega $ of the model. The parameter $\omega $,
having the physical dimensions of ${\rm length}^{-2}$, is
constrained by the perihelion precession of the planet Mercury to
a value of $\omega \geq 7\times 10^{-16}/M_{\odot }^2=3.212\times
10^{-26}$ cm$^{-2}$. The deflection angle of the light rays by the
Sun can be fully explained in Ho\v{r}ava-Lifshitz gravity with the
parameter $\omega $ having the value $\omega \geq 10^{-15}/M_{\odot
}^2=4.5899\times 10^{-26}\;{\rm cm}^{-2}$, while the radar echo delay experiment
suggests a value of $\omega \geq 2\times 10^{-15}/M_{\odot
}^2=9.1798\times 10^{-26}$ cm$^{-2}$. Tentatively, and in order to provide a numerical value that could be used in practical calculations, from these values we can
give an estimate of $\omega $ as
\begin{equation}
\omega =\left(5.660\pm3.1\right)\times 10^{-26}\;{\rm cm}^{-2}.
\end{equation}

The standard deviation in our determination of $\omega $ is
$3.2\times 10^{-26}$ cm$^{-2}$, the variance is $9.7\times
10^{-26}$ cm$^{_2}$, and the median of the determined values is
$4.5\times 10^{-26}$ cm$^{_2}$. It is interesting to note that the
values of $\omega $ obtained from the study of the light
deflection by the Sun and of the radar echo delay experiment are
extremely close, while the perihelion precession of the planet
Mercury provides a smaller value. Unfortunately, the observational
data on the perihelion precession are strongly affected by the
solar oblateness, whose value is poorly known \cite{Ca83, RoRo97}. We have also neglected the solar Lense-Thirring effect \cite{iorio3}, as well as the effects of the asteroids.
 Even so, by
taking into account the smallness of the parameter $\omega $, it
follows that there is a very good agreement between the numerical
values for $\omega $ obtained from these three Solar System Tests. On the other hand, it is important to note that the light deflection and the radar echo delay observations, which are very similar from a physical point of view, do not give intersecting intervals for $\omega _0$, since the value of $\omega _0$ obtained from light deflection is in the range $\omega _0\in
\left(1.1\times10^{-15},1.3\times 10^{-15}\right)$, while from the radar echo delay we obtain $\omega _0\in\left(2.0199\times 10^{-15},2.2000\times
10^{-15}\right)$. The values obtained from the light deflection by the Sun are systematically smaller, by a factor of around two, as compared to the values obtained from radar echo delay observations. This non-intersecting range of values could be explained by the differences in the observational errors for the two effects. While the observational error in light deflection is around 0.0017, the corresponding error in the radar echo delay observations is of the order of  $ 10^{-5}-10^{-6}$. The very small error of the radar echo delay allows a very precise determination of the value of $\omega _0$. Another possibility for this discrepancy may be related to some intrinsic properties of the model, like, for example, the fact that in Ho\v{r}ava-Lifshitz gravity there is no full diffeomorphism invariance of the Hamiltonian formalism.

In the weak-field and slow-motion approximation, the corrections to the third Kepler law of a test particle in the Kehagias and Sfetsos black hole geometry were obtained in \cite{iorio}. The corrections were compared to the phenomenologically determined orbital period of the transiting extrasolar planet HD209458b Osiris. The order-of-magnitude lower bound on the parameter $\omega _0$ obtained from this study is $\omega_0 \geq 1.4\times 10^{-18}$, as compared to the value $\omega _0\geq 7\times 10^{-16}$ obtained in the present paper. Tighter constraints are established by the inner planets for
which $\omega _0\geq 10^{-15}- 10^{-12}$ \cite{iorio}. However, in order to obtain a better precision from these data, a full general relativistic study is needed, as well as a significant improvement in the determination of the values of the orbital periods of the exoplanets.

Thus, the gravitational dynamics of the KS solution is determined
by the free parameter $\omega $. In order to explain the
observational effects in the Solar System, $\omega $ must have an
extremely small value, of the order of a few $10^{-26}$ cm$^{-2}$.
Therefore the explanation of the classical tests of GR must
require a very precise fine tuning of this constant at the level
of the Solar System. It is also very important for future
observations to determine if $\omega $ is a local quantity or a
universal constant.
By assuming that $\omega $ is a universal constant, its smallness
suggests the possibility that it may have a microscopic origin.

In conclusion, the study of the classical tests of general
relativity provide a very powerful method for constraining the
allowed parameter space of the Ho\v{r}ava-Lifshitz gravity
solutions, and to provide a deeper insight into the physical
nature and properties of the corresponding spacetime metrics.
Therefore, this opens the possibility of testing
Ho\v{r}ava-Lifshitz gravity by using astronomical and
astrophysical observations at the Solar System scale \cite{iorio}. Of course,
this analysis requires developing general methods for the high
precision study of the classical tests in arbitrary spherically
symmetric spacetimes. In the present paper we have provided some
basic theoretical tools necessary for the in depth comparison of
the predictions of the Ho\v{r}ava-Lifshitz gravity model with the
observational/experimental results.

\section*{Acknowledgments}

We would like to thank the two anonymous referees for comments and suggestions that helped us to significantly improve our manuscript.  The work of T. H. was supported by the General Research Fund grant
number HKU 701808P of the government of the Hong Kong Special
Administrative Region. FSNL acknowledges partial financial support of the Funda\c{c}\~{a}o para a Ci\^{e}ncia e Tecnologia through the grants PTDC/FIS/102742/2008 and CERN/FP/109381/2009.

\end{document}